\documentclass[preprint,double-spaced,aps,apssymb]{revtex4}
\usepackage{amsmath,amssymb}
\usepackage{graphicx}% Include figure files
\usepackage{dcolumn}% Align table columns on decimal point
\usepackage{bm}% bold math
\newcommand{\ber}{\begin{eqnarray}}
\newcommand{\eer}{\end{eqnarray}}
\newcommand{\bea}{\begin{equation}}
\newcommand{\eea}{\end{equation}}

\begin{document}

\title{Local structures in the resistive state of a one dimensional superconductor}

\author{A. Bhattacharyay}
\email{a.bhattacharyay@iiserpune.ac.in}
 \affiliation{Indian Institute of Science Education and Research, Pune, India}

\date{\today}

\begin{abstract}
In a one dimensional superconductor where current driven phase transitions occur between superconducting and normal phases, both the phases coexist in a metastable regime over a wide range of current near the critical current $j_c$. A lot of spatio-temporal localized forms of the the competing phases have been identified in this so called resistive regime. In this paper we present the relation that selects the closed conjugate form for the other order parameter when that of the one of two competing states is known. Our main observation is that, the free energy surface the system remains predominantly bound to in the resistive regime is dominated by the quadratic term of the phenomenological Ginzburg-Landau potential.
\end{abstract}
\pacs{74.20.De, 89.75.Kd, 85.25.Am}% PACS, the Physics and Astronomy
\maketitle
One dimensional (1D) superconductor has its width smaller than the Ginzburg-Landau (G-L) coherence length $\xi(T)$ (average separation between electrons in a cooper pair) where T is the temperature of the system. Near the critical temperature $T_c$ of the superconducting (SC) to normal (N) phase transition, the coherence length $\xi(T)$ is quite large and a finite width channel or wire of superconducting material can be treated as a 1D system. Such a system has the pure SC state (characterized by having no Ohmic voltage drop along the wire despite having current through it) globally stable below a critical current strength $j_c$ (G-L critical current) at $T<T_c$. At a current strength $j>j_c$, being induced by an applied voltage across the length of the wire, the SC state coexists with the N state up to an upper limiting current $j_2$. Beyond $j_2$ the N state is globally stable. Within the limits $j_c<j<j_2$, the SC state is metastable and gives up locally to the N state by occasional formation of phase slip centers (PSCs) \cite{lan,mac,kop}. Phase sleep centers are the points in space where the amplitude of the SC order parameter becomes zero. By the formation of phase sleep centers the complex order parameter of the SC state loses a turn along the length of the wire and moves on to a state characterized by a larger wavelength and lower free energy. In this way, by steps, the SC state transforms to an N state \cite{lan,mac,kop,grp1}. Various localized forms of the SC amplitude have mostly been numerically observed to exist in this regime.
\par
When one comes down from the N state to the SC state by slowly lowering the current j, one also observes a very important role played by localized instabilities of size beyond a limit called critical nucleus. The N state is not globally stable in this regime and loses stability to a localized SC state of size bigger than the critical nucleus whereas it remains stable against smaller fluctuations \cite{kop,wat}. These critical nuclei have been numerically investigated assuming them to stem from the non-linearity of the system. The size of this critical nucleus vanishes at $j_c$ where the SC state can take over from the N state due to infinitesimal fluctuations and it increases towards $j_2$ indicating the transition to SC state to be more difficult and energetically not favoured since large fluctuations should be rarely occurring. Finite sized localized structures of the SC order parameter and the chemical potential $\mu$ (N state order parameter) are therefore the most important entities in this intermediate regime. By nucleation of these, the system shows a slow transition from one state to the other. One way of understanding the spontaneous generation of the localized structures focuses on the physics at the microscopic scales, break down of a cooper pair induced by the applied electric field, subsequent acceleration of excited electrons, its relaxation and repairing etc (see ref.\cite{kop} and references therein). On the other hand, identifying the set of localized solutions, static as well as dynamic, allowed by the equations that govern the relaxation of the system in this regime is something of immense importance and its this latter objective that we will be discussing in the present context. 
\par
Recently, a lot of work are being done in 1D superconductors which show finite resistance at temperatures quite close to absolute zero due to quantum fluctuations (see ref.\cite{aru} and references therein). Although, the mean field theory cannot account for the quantum fluctuations but the mechanism on the basis of the energetics are the same where the role of thermal fluctuations in driving the system from one metastable state to the other is played by quantum fluctuations. In the meanfield regime, the localized and transient modulations of the morphology of the relevant global order parameters (PSCs and critical nuclei) are due to excursion of the trajectory from one local minima of relevant free energy to another via one of many possible routes over the free energy barrier. However, in the quantum regime, the wave function tunnels through the barrier to the lower free energy configuration with qualitatively similar manifestation of quantum PSCs and there are a considerable similarity at the level of structural aspects of the order parameter variations in these two different regimes of qualitatively different origin of fluctuations. Some structural properties of the fluctuations could be to a great extent ascribed to the symmetries and the form of the free energy of the system irrespective of the order parameter is mean field or not. In the present work, we are going to show that, given the complex G-L free energy of the system there are two length scales that dominate the proceedings. The smaller length scale is that of the wavelength of the SC order parameter. If we isolate the dynamics of the system at a length scale much bigger than the smaller length scale we readily get a relation that guides the morphology of fluctuations in competing order parameters and gives us closed analytic form for a host of such localized spatial structures.
\par
In the present work, we are analysing the local structures using the tools of pattern formation in dynamical systems. We have separated the dynamics on two widely varying scales to use the technique of multiple scale perturbations generally applied with the help of a linearized system at the vicinity of an instability threshold. The typical structure of the non-linearity of the complex G-L equation, in its simplest form, that captures the dynamics of a 1D SC system in its resistive state (metastable regime where SC and N orders coexist), enables us to consider the dynamics to be essentially linear on the smaller scales. The invariance of the system under a global translation in phase (order parameter multiplied by $e^{i\phi}$) would allow us to linearise the dynamics at the small scales by considering the amplitude to vary at larger scales, and that simplifies the problem considerably. The slow dynamics of the amplitude and the wave number of the system separates out as a solvability condition for the existence of the small scale results and are looked at to identify the local structures in them. From the amplitude equation we will derive a relationship between the SC and N order parameters which manifests almost every observed type of conjugate forms of them. On the basis of this relation, we present here a variety of closed analytic forms of the SC localized states and corresponding chemical potential ($\mu$) whose look alikes have numerically been shown before. We capture uniformly moving fronts pushing away localized SC/N states  from expanding global N/SC phases on the basis of same equations. We correlate our observations with the already existent theoretical understanding of the fluctuation induced appearance of PSCs to realize that the trajectory of the transient system is mostly bound to the quadratic term dominated portion of the free energy surface and that underlies the success of the present linear analysis.
\par
The phenomenological complex G-L (CGL) model - in its simplest form - that captures the dynamics of the a 1D superconductors in general is the following
\ber\nonumber
u(\psi_t+i\mu\psi)&=&\psi_{xx} + (1-\vert\psi\vert^2)\psi \\
j&=&Im(\psi^*\bigtriangledown\psi)-\mu_x.
\eer          
In the above model, $\psi$ is the superconducting order parameter which is complex valued. The system being 1D, the $x$ is distance along the wire from some arbitrary origin. The $\mu$ is electrochemical potential which is also the order parameter of the N phase of the system. The $j$ is the current (density) through the superconducting wire and the parameter $u$ is the ratio of the order parameter relaxation time to the current relaxation time. In what follows we will treat $u$ as a constant, since, in the present context we are not going to investigate the competing dynamics of various localized states rather would identify the closed form analytic expressions for them. The typical value of $u$ is 12 in the strong depairing limit whereas $u=5.79$ for weak depairing \cite{scm}. The suffix $t$ and $x$ of $\psi$ indicate of the partial derivatives of the SC order parameter with respect to them. 
\par
Eq.1 has two stationary solutions. (1) $\psi \equiv 0 $, $\mu = -xj$ (for constant $j$) which is the normal state and (2) $\psi=Ae^{iqx}$, $q^2=1-A^2$, $j=A^2q$, which is the superconducting state when $\mu\equiv 0$ \cite{kra}.  As has already been stated, the resistive state, in which the SC and the N phases coexist, falls within the current limits $j_c<j<j_2$ \cite{kop, kra}. An extensive numerical analysis by Kramer et al in this region showed localized SC phases and corresponding $\mu$ profiles, static and dynamic PSCs \cite{kra} etc. In this same resistive regime, Watts-Tobin {\it et al.} \cite{wat} numerically studied the so called critical nuclei by the formation of which (localized structures) the N phase loses stability to the SC phase. All these extensive numerical studies have been on the basis of the CGL model as mentioned above or on its variants and that is why it has been chosen for analytic investigations in the present context.       
\par
In the resistive state, let us consider the order parameter $\psi$ to have the form $\psi=A(X,\tau)[\psi^0+\epsilon\psi^1]$ where the amplitude of the order parameter $A(X,\tau)$ is a function of slow space and time scales $X$ and $\tau$ respectively. Let us also consider the N phase order parameter $\mu$ to be $O(\epsilon)$. We are interested in getting the dynamics of $A(X,\tau)$ in such a situation. Since, the dynamics of the system in this resistive state is slow, the only time scale under effective consideration here is the slow one $t = \epsilon \tau$. Multiple length scales as $x=x+\epsilon X$ are used to capture the slow evolution of the amplitude and wave vector $q(X,\tau)=-i\partial_x\psi$ of the otherwise fast SC order parameter in space. On consideration that the scales $x$ and $X$ are widely separated, the model (Eq.1) is rewritten as,
\ber\nonumber
u(\psi_t+i\mu\psi)&=&\psi_{xx} + (1-A^2)\psi \\
j&=&Im(\psi^*\bigtriangledown\psi)-\mu_x.
\eer  
where $A^2$ will be treated as a constant since $\psi$ is a much faster varying function than its amplitude. This renders the model linear without changing any of the steady state properties of the SC and the N phase solutions of it. The standard tools of singular perturbations can now be brought to bear with this. Expanding up to $O(\epsilon)$ we get
\ber\nonumber
&u&\left(\epsilon\frac{\partial A(X,\tau)}{\partial \tau}[\psi^0+\epsilon\psi^1]+\epsilon A(X,\tau)\frac{\partial\psi^0}{\partial \tau}\right)\\\nonumber & =& A(X,\tau)\left[ \frac{\partial^2\psi^0}{\partial x^2} + \epsilon\frac{\partial^2\psi^1}{\partial x^2} \right] \epsilon 2\frac{\partial A(X,\tau)}{\partial X}\left[ \frac{\partial\psi^0}{\partial x} + \epsilon\frac{\partial\psi^1}{\partial x} \right]\\ &+& \epsilon 2A(X,\tau)\frac{\partial^2\psi^0}{\partial X\partial x} + (1-A^2)A(X,\tau)[\psi^0+\epsilon\psi^1].
\eer
\par
Thus, at $O(1)$,
\bea
\psi_{xx}^0 +(1-A^2)\psi^0=0,
\eea 
is a zero eigenvalue equation and the solution is of the form $\psi^0=A(X,\tau)e^{iq(X,\tau)x}$. In the next order ($O(\epsilon)$),
\ber
\nonumber
&&\psi_{xx}^1 + (1-A^2)\psi^1=( u\frac{\partial A(X,\tau)}{\partial \tau} + iu\mu A(X,\tau)\\\nonumber &-& i2q(X,\tau)\frac{\partial A(X,\tau)}{\partial X} +iuxA(X,\tau)\frac{\partial q(X,\tau)}{\partial \tau}\\ &+& (2xq(X,\tau)-2i)A(X,\tau)\frac{\partial q(X,\tau)}{\partial X})\psi^0
\eer
The solvability at this order requires the right hand side of the Eq.5 vanish. Considering the evolution of the wave vector to be independent of the amplitude dynamics we separate the equations for the amplitude and the wave vector as
\ber
\frac{\partial A}{\partial t} +i\epsilon\mu A - i\frac{2q}{u}\frac{\partial A}{\partial x}&=& 0\\
iux\frac{\partial q}{\partial t} +2(xq-i)\frac{\partial q}{\partial x}&=& 0.
\eer
In Eq.6 and 7 the smaller scales have been restored which effectively introduces an $\epsilon$ factor to the middle term of the Eq.6. 
\par
The fixed points of the Eq.7 are $q=q_0$ (constant) and $q=i/x$. Considering $q=q_0$, the stationary amplitude $A(x)$ of the SC phase or the effective SC order parameter is related to the stationary N phase order parameter $\mu(x)$ by the relation
\bea
\frac{d}{dx}\ln{A(x)} = \frac{\epsilon u}{2q_0}\mu (x).
\eea
Eq.8 is the sought after central relationship which correlates the spatial profiles of the two competing phases in the resistive state and would manifest a number of interesting local structures. According to Eq.8 the N phase order parameter is identically zero for a constant amplitude SC phase which is perfectly consistent with what one expects for a pure SC state. The numerical investigation of Kramer {\it et al.} \cite{kra}, in this resistive regime, has shown a Gaussian looking localized profile of SC amplitude associated with an almost linear form of the $\mu$ passing through the origin. Let us proceed to identify the local structures in $A(x)$ and $\mu$ being guided by these numerical observations. Consider a localize form of $A(x)=A_0e^{-a_0x^2}$, using Eq.8 one immediately gets
\bea
\mu=-\frac{4a_0q_0}{\epsilon u}x,
\eea
a linear profile of the N phase order parameter passing through the origin. By the use of Eq.8, thus, one can easily analytically obtain the closed forms of localized solution of CGL model the analogues of which were numerically shown before. It comes out that a stationary, Gaussian, localized SC phase coexists with a stationary global N phase with a normal current $j_N=-\frac{4a_0q_0}{\epsilon u}$. This form is a special case of the general solution $A(x)=e^{-f(x)}$ where f(x) is any useful representative function that characterizes the SC amplitude and the corresponding $\mu$ is proportional to $-f^\prime (x)$. This particular form has already been shown associated with dynamic PSCs \cite{ari}.
\par
Consider $A(x)=A_0 sech(\alpha x)$, the N phase order parameter comes out as
\bea
\mu=-A_0\frac{2q\alpha}{\epsilon\mu}\tanh{(\alpha x)}.
\eea  
Thus, a hyperbolic secant form of the SC phase should be observed at the point where the $\mu$ front, that connects two N phases, characterized by $\mu_{\pm}=\pm A_0\frac{2q\alpha}{\epsilon\mu}$, passes through zero or the N phase locally vanishes. Such a localized form is more expected to be seen on a smaller system size, where, the constant potentials at the end of the wire are represented by $\pm\mu$. Another interesting localized forms of $A(x)$ is a $tanh(\vert x\vert)$ form which represents a static phase slip center of the SC phase. Basically, this form represents a hole connecting two SC phases of phase difference $2\pi$ along the wire. At the origin, where the PSC is sitting the corresponding $\mu(x)\propto [1/tanh(x)-tanh(x)]$ will have a singularity. The $\mu$ would diverge at the point where $A(x)$ is zero keeping the symmetries intact. One can also get $\mu(x)$ in a form which is a combination of the relations 9 and 10 by taking $A(x)=e^{-x^2}cosh(x)$. This is a bounded form of the SC order parameter which vanishes at large distances. Eq.8 gives us the corresponding form of $\mu$ as $\mu\propto -2x + tanh(x)$. Thus, making use of the relation 8, one can easily uncover a host of compatible closed solutions of the two competing orders in the resistive state of a 1D superconductor to which the CGL applies.
\par
As another special case, consider a front of the form $A(x)\propto e^x/(1+e^x)$ that separates an SC phase at $+\infty$ to a non SC phase at $-\infty$. Corresponding $\mu$ profile as obtained from Eq.8 is $\mu(x)\propto 1-e^x/(1+e^x)$ which connects an N phase at $-\infty$ to a non N phase at $+\infty$. The front, under consideration, actually separates a N state from the SC state in the resistive regime. Taking the $q_0=\alpha +i\beta$ a complex number we can readily go to a co-moving frame ($x=x-vt$) of velocity $v=2\beta/u$ keeping Eq.8 unaltered. Here $\beta$ is the spatial damping rate of the SC phase. Interestingly, for a positive $\beta$ i.e. when the SC phase is damped at large distances, the $A(x)=e^x/(1+e^x)$ front will move towards the SC phase from the N phase pushing the localized SC phase towards one end of the wire. With $\beta$ negative i.e. when the SC phase is growing over space, the front moves from the SC region towards the N region. This method of shifting to a co-moving frame can be applied to any of the above mentioned solutions because it preserves the structure of Eq.8. Basically the spatial damping and the growth on the other sides of the origin break the symmetry about it. This symmetry breaking will result in peaking up of a direction for the uniform motion of the fronts. So, practically all the variants of the local structures discussed so far can exist in a uniformly moving state, of course with some associated distortions in shape due to symmetry breaking about the origin.
\par
To understand the implications of the existence of above mentioned localized forms on the stability of the SC phase in the resistive regime, consider perturbing the stationary (or uniformly moving) $A_0(x,t)$, $q_0$, where $A_0(x,t)$ is related to $\mu_0$ by Eq.8, with infinitesimal perturbations $\delta A$ and $\delta q_0$. One can readily linearise Eq.6 and 7 in a general form, for all of the above mentioned fixed points (local structures) of these equations, to get
\bea
\frac{\partial}{\partial t} \begin{pmatrix} \delta A \\ \delta q \end{pmatrix} = \begin{pmatrix} {-(i \epsilon \mu_0+\frac{2kq_0}{u})} & {i\frac{2}{u}\frac{\partial A}{\partial x}} \\
0 & {ik(\frac{2}{ux}+i\frac{2q_0}{u})} \end{pmatrix} \begin{pmatrix} \delta A \\ \delta q \end{pmatrix},
\eea
where $k$ is the characteristic wave number of the perturbations. This immediately gives the growth rate of the perturbations as $\lambda_{1,2}=-(i\epsilon\mu_0+\frac{2kq_0}{u}),(i\frac{2k}{ux}-\frac{2kq_0}{u})$. The real part of the growth rate $\lambda_{r}=-\frac{2kq_0}{u}$ is positive for opposite signs of $k$ and $q_0$. This is consistent with the way the SC state is observed to reduce free energy by gradually moving to higher wavelengths. What interesting about this linear stability analysis is that it directly shows in the presence of solutions (fixed points) of the Eq.8 the SC state is locally unstable to perturbations that would enhance the wavelength of the state. The constant amplitude SC phase is known to be locally stable and globally unstable in the resistive regime. The allowed amplitude modulations - the exact solutions of Eq.8 - facilitate the SC to N transition by making the SC state locally unstable.  
\par
The analysis we have gone through so far strongly indicates that the nonlinear part of the dynamical system has nothing to do with the selection of the functions dominating the scenario in the resistive state. This is not surprising though. Let us have a look at the expression of the free energy (F) for a wire of length L corresponding to the dynamics we are dealing with 
\bea
F=\sigma L[(q^2-1)A^2+\frac{A^4}{2}]
\eea
where $\sigma$ is the cross section of the wire. We can clearly see from the above expression that F has the lowest value corresponding to the wave number $q=0$ for all A. Fig.1 shows how the free energy is changing at the $|q|$ values 2, 1 and 0. When $|q|\geq 1$ the system has a single minimum at $A=0$, as soon as $|q| $ goes less than 1 the system starts developing two other minima about the former and these minima reaches their lowest value at q=0.
\par
To understand why the system remains bound to the quadratic term dominated surface of the free energy we should recall the qualitative picture of the PSC formation as understood in the light of the classic theory put forward by Little \cite{lit}, Langer-Ambegaokar (LA) and McCumber-Halperin (MH) \cite{lan,mac}. LA developed the theory on the basis of an idea put forward by Little that the superconducting order parameter amplitude vanishes locally to help the spiral (order parameter is complex) lose a turn and the subsequent relaxation of the turns of the spiraling superconducting order parameter moves from a state of wave number $q_0$ to $q_0-2\pi/L$ where L is the system size with periodic boundary conditions taken into consideration. The LA theory subsequently was improved by MH by taking into consideration the dynamics and number of available states along the wire. Anyway, this is essentially the physical understanding of how PSCs from and because of this mechanism a single PSC at a particular time can only allow the system to lose one turn i.e. wave number by the amount $2\pi/L$. 
\par
At this stage we refer to Fig.2 where a schematic diagram of the free energy has been plotted for the states with wave numbers $q_0$ (upper branch) and $q_0-2\pi/L$ (lower branch). Since, so long as the amplitude does not exactly become zero at a point to form the PSC, the wave number of the SC phase remains necessarily unaltered (number of turns are unaltered), during this process of the formation of the PSC the system will move uphill (induced by thermal fluctuations) from A to O where the amplitude actually becomes zero. Now, being at O the amplitude loses one turn and eventually takes the lower branch to come down to the minima at B corresponding to the wave number $q_0-2\pi/L$. So, through out this transition the wave number remains fixed either at $q_0$ or $q_0-2\pi/L$ which makes the the system to traverse the free energy surfaces dominated by the quadratic terms of the GL free energy expression and thus any functional forms of amplitude consistent to this excursion be necessarily found out from the linear part of the dynamical theory. In fact, in our present analysis we have exactly done the same thing by keeping the wave number fixed and considering only the variations of the amplitude of the SC phase and the $\mu$. Although, this problem is a classic one (almost 40 years old), yet, to our knowledge nobody had previously illustrated this aspect of the system that remaining bound to the quadratic term dominated free energy surface the dynamics is essentially linear. However, the present analysis clearly shows that a host of such localized amplitude modulations of the resistive state can be captured on the basis of the linear dynamics only and that on quite evident reasons.
\par
To conclude, we have arrived at a relation between the two competing order parameters, in the resistive state of a 1D superconductor, which is central to identify the compatible forms of spatially modulated $A$ and $\mu$. We utilized the linear CGL equation on the basis of the consideration of dynamics at widely separated scales to arrive at the linear amplitude equation for the SC order parameter. The invariance of CGL under arbitrary translation of global phase allows such separation of scales. We have captured a lot of local structures, uniformly moving fronts etc. for the competing SC and N phases and it comes out that, the large family of these local structures formation is not non-linearity driven. We have done a stability analysis of the amplitude modulated SC state to discover that it becomes locally unstable in the presence of these modulations. Thus, the role of this family of amplitude modulations in the resistive regime is also understood as a facilitators to the phase transitions which historically is the basic understanding of the theory put forward by Langer-Ambegaokar and McCumber-Halpering \cite{lan,mac}. All these local fluctuations are consequences of the motion of the trajectory over similar free energy barriers between local minima and that makes it not surprising to have such a unifying relation resulting from the CGLE dynamics which captures a whole lot of localized modulations of the order parameters or critical nuclei on the same footing.

\newpage

\begin{figure}
\begin{center}
\includegraphics[width=12cm,angle=0]{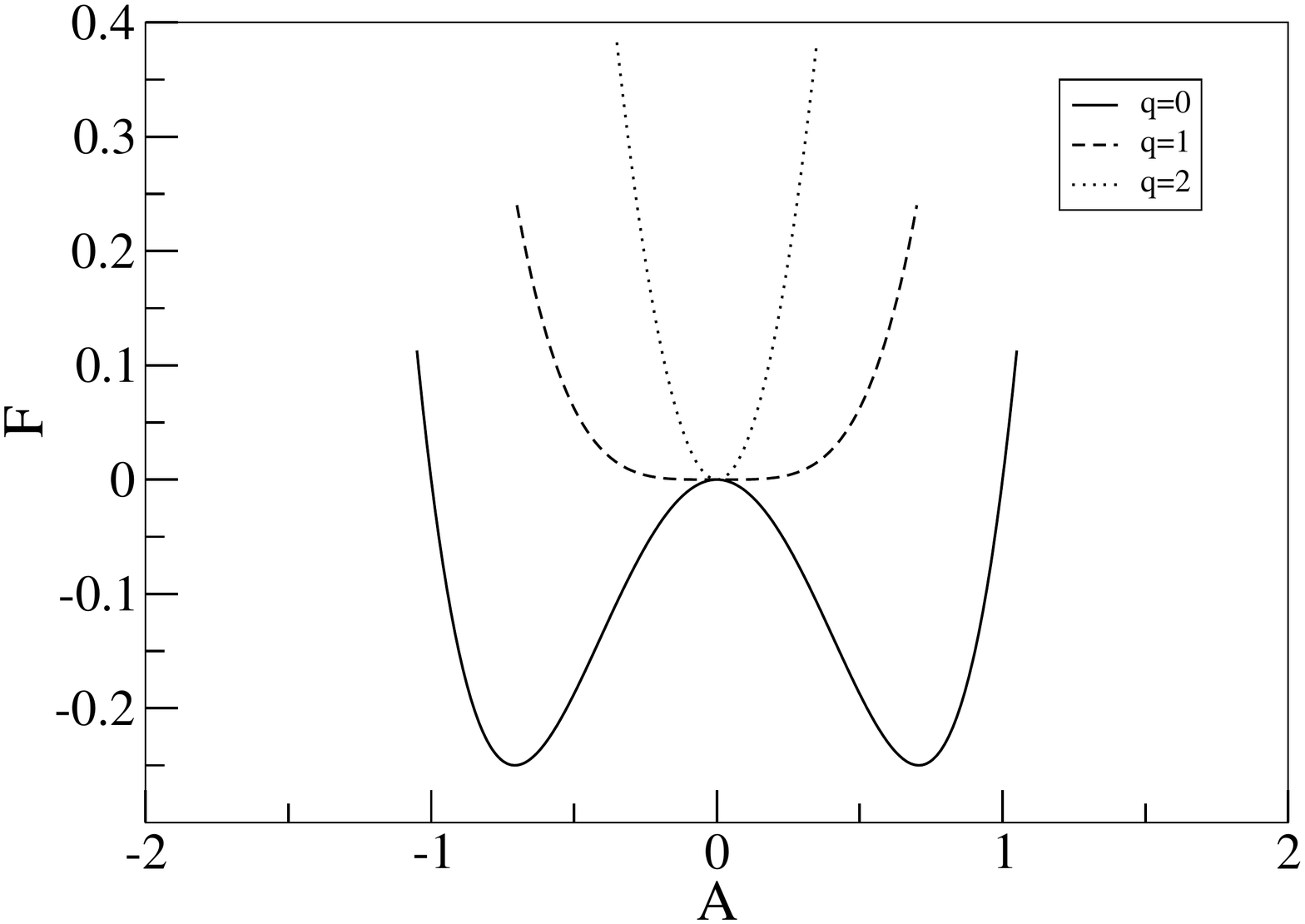}
\caption[Figure 1]{Free energy plots for the values of the wave number 0, 1 and 2.}
\end{center}
%\end{minipage}
\end{figure}

\begin{figure}
\begin{center}
\includegraphics[width=12cm,angle=0]{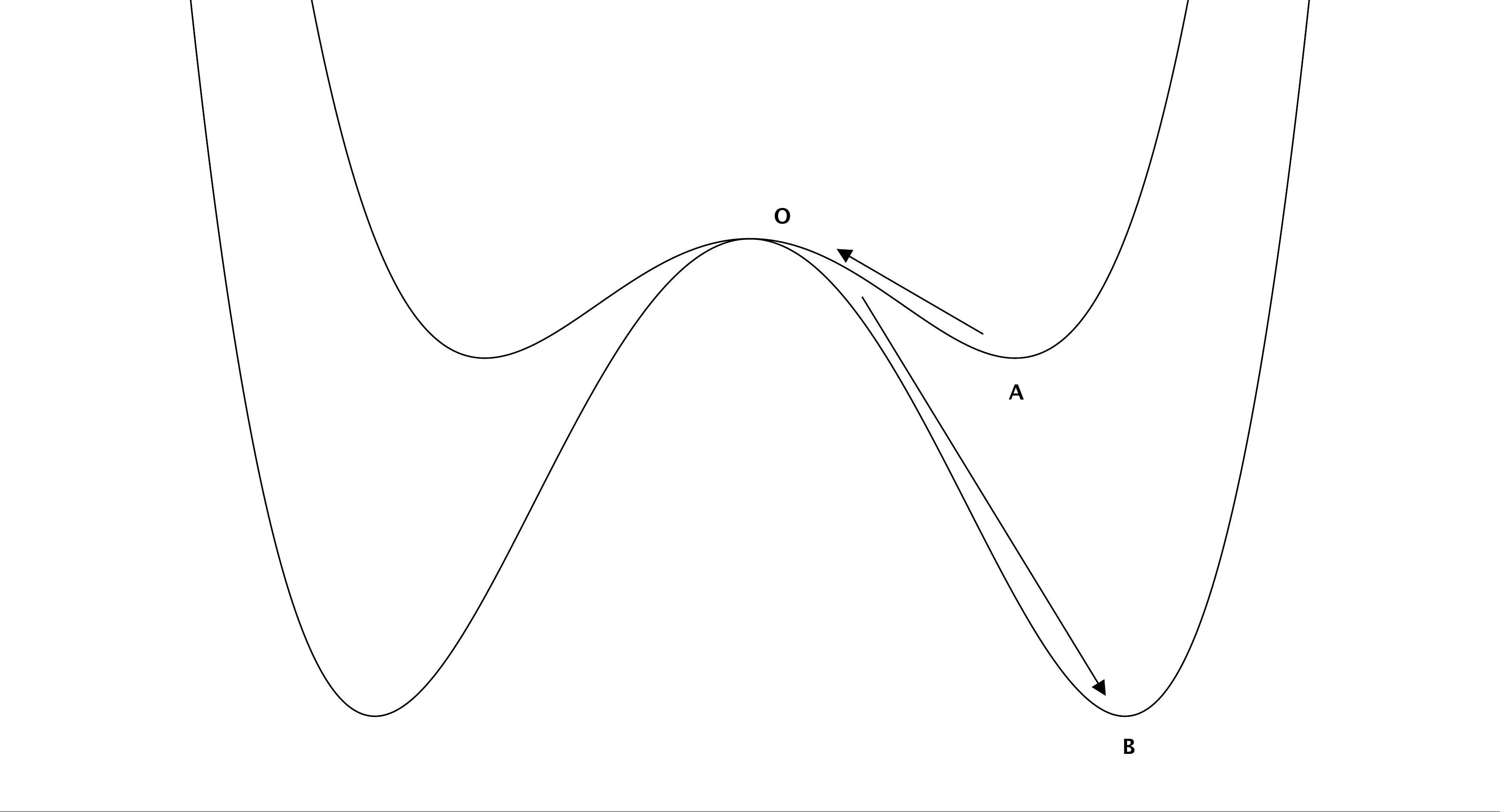}
\caption[Figure 2]{Schematic plot of free energy surfaces to understand how the system moves along those at the time of formation of a PSC.}
\end{center}
%\end{minipage}
\end{figure}


\newpage
\begin{thebibliography}
\noindent\bibitem {lan} Langer J S and Ambegaokar V, Phys. Rev. {\bf 164} 498 (1967)
\noindent\bibitem {mac} McCumber D E and Halperin B I, Phys. Rev. B {\bf 1} 1054 (1970)
\noindent\bibitem {kop} Ivlev B I and Kopnin N B, Adv. Phys. {\bf 33} 47 (1984)
\noindent\bibitem {grp1} Meyer J D and Minnigerode G V, Phys. Lett. A {\bf 38}, 529 (1972); Skocpol N J, Beasley M R and Tinkham M, J low-temp. Phys. {\bf 16}, 145 (1974); Skocpol N J, Beasley M R and Tinkham M, J Appl. Phys. {\bf 45}, 4054 (1974); Dmitriev V M and Khristenko E V, Fiz. Nizk. Temp. {\bf 3}, 1210 (1977); Smith L N, J low-temp. Phys. {\bf 38}, 553 (1980)
\noindent\bibitem {wat} Watts-Tobin R J, Kr\"ahanb\"uhl Y and Kramer L, J. Low-temp. Phys. {\bf 42}, 459 (1981)
\noindent\bibitem {aru} K Yu Arutyunov, D S Golubev and A D Zaikin, Phys. Rep. {\bf 464}, 1 (2008)
%\noindent\bibitem {zha} Zharov A, Lopatin A, Koshelev A E and Vinokur V M 2007 Phys. Rev. Lett. {\bf 98} 197005
\noindent\bibitem {scm} Schmid A 1966 Phys. Kondens Mater {\bf 5} 302 
\noindent\bibitem {kra} Kramer L and Baratoff A 1977 Phys. Rev. Lett. {\bf 38} 518
%\noindent\bibitem {rub} Rubinstein J, Sternberg P and Ma Q 2007 Phys. Rev. Lett. {\bf 99} 167003
\noindent\bibitem {ari} Bhattacharyay A, J. Phys. A: Math. Theor. {\bf 41}, 112001 (2008)
\noindent\bibitem {lit} Little W A, Phys. Rev. {\bf 156}, 396 (1967)
\end{thebibliography}
\end{document}